\documentclass[12pt]{article}
\usepackage{epsf}
\def\1ad{\mbox{\normalsize $^1$}}
\def\2ad{\mbox{\normalsize $^2$}}
\def\3ad{\mbox{\normalsize $^3$}}
\def\4ad{\mbox{\normalsize $^4$}}
\def\5ad{\mbox{\normalsize $^5$}}
\def\6ad{\mbox{\normalsize $^6$}}
\def\7ad{\mbox{\normalsize $^7$}}
\def\8ad{\mbox{\normalsize $^8$}}

\def\beq{\begin{equation}}                     %
\def\eeq{\end{equation}}                       %
\def\bea{\begin{eqnarray}}                     %         %
\def\eea{\end{eqnarray}}                       %       %
\begin{document}

\newcommand{\sheptitle}
{Topology Change
 and Unitarity in  Quantum Black Hole  Dynamics
}
\newcommand{\shepauthora}
{{\sc
 J.L.F.~Barb\'on}
}

\newcommand{\shepaddressa}
{\sl
Instituto de F\'{\i}sica Te\'orica  IFT UAM/CSIC \\
 Facultad de Ciencias C-XVI \\
C.U. Cantoblanco, E-28049 Madrid, Spain\\
{\tt jose.barbon@uam.es}}

\newcommand{\shepauthorb}
{\sc
E.~Rabinovici}

\newcommand{\shepaddressb}
{\sl
Racah Institute of Physics, The Hebrew University \\ Jerusalem 91904, Israel
 \\
{\tt eliezer@vms.huji.ac.il}}
\newcommand{\shepabstract}
{
We discuss to what extent semiclassical topology change is capable of
restoring unitarity in the relaxation of perturbations of eternal black holes
in thermal equilibrium. The Poincar\'e recurrences required by unitarity
are not correctly reproduced in detail, but their effect on infinite
time-averages can be mimicked by these semiclassical topological
 fluctuations. We also discuss the possible implications of these
 facts to the question of unitarity of the black hole S-matrix~\footnote{
Talk
delivered  by JLFB
at ERES2004 ``Beyond General Relativity". Miraflores de la Sierra, Madrid,
september 2004. }.
}

\begin{titlepage}
\begin{flushright}
{IFT-UAM/CSIC 05-019\\
{\tt hep-th/0503144}}

\end{flushright}
\vspace{0.5in}
\begin{center}
{\large{\bf \sheptitle}}
\bigskip\bigskip \\ \shepauthora \\ \mbox{} \\ {\it \shepaddressa} \\
\vspace{0.2in}
%\vspace{1in}
\bigskip\bigskip  \shepauthorb \\ \mbox{} \\ {\it \shepaddressb} \\
%\vspace{0.5in}
\vspace{0.5in}

{\bf Abstract} \bigskip \end{center} \setcounter{page}{0}
 \shepabstract
\vspace{1.5in}
\begin{flushleft}
%CERN-TH/2004-059\\
%March 2001
\today
\end{flushleft}
\end{titlepage}

\newpage

%%%%%%%%%%%%%%%%%%%%%%%%%%%%%%%%%%%%%%%%%%%%%%%%%%%%%%%%%%%%%%%%%%%%%%

\section{Introduction}

\noindent

\setcounter{equation}{0}

The AdS/CFT correspondence \cite{rads} 
provides a nonperturbative model of quantum gravity in which black holes
seem to form and evaporate as standard unitary processes in quantum mechanics.
 In this construction,
 quantum gravity
in a $d$-dimensional  asymptotically Anti-de Sitter spacetime
 (AdS) of curvature radius $R$ is {\it defined}
in terms
of  a conformal field theory (CFT) on a spatial sphere ${\bf S}^{d-2}$ of
radius $R$.
The effective  expansion parameter on the gravity side $1/N^2 \sim G_{\rm N}/R^{d-2}$, maps to
an appropriate large $N$ limit of the CFT. For example, for two-dimensional CFT's
$N^2$ is proportional to the central charge. When the
 CFT is a
gauge theory and the AdS side is a string theory, $N$
is the rank of the gauge group,  and the string perturbative  expansion in powers
of $g_s \sim 1/N$
is identified with 't Hooft's $1/N$ expansion in the gauge theory side.

According to this definition, the formation and evaporation of small black holes
with Schwarschild radius  $R_S \ll R$,
 should be described by a unitary process in terms of
the CFT Hamiltonian. Thus, there is no room for violations of coherence,
independently of the manner in which the process is encoded in the CFT. 
 Unfortunately, the CFT states corresponding to small black holes
are hard to describe, and it remains a challenge to put the finger on the precise
error in the standard  semiclassical analysis \cite{rsdollar} in that case.

\begin{figure}[hp] 
%\hspace*{0.1in}
\begin{center}
\epsfysize=1.5in
\epsffile{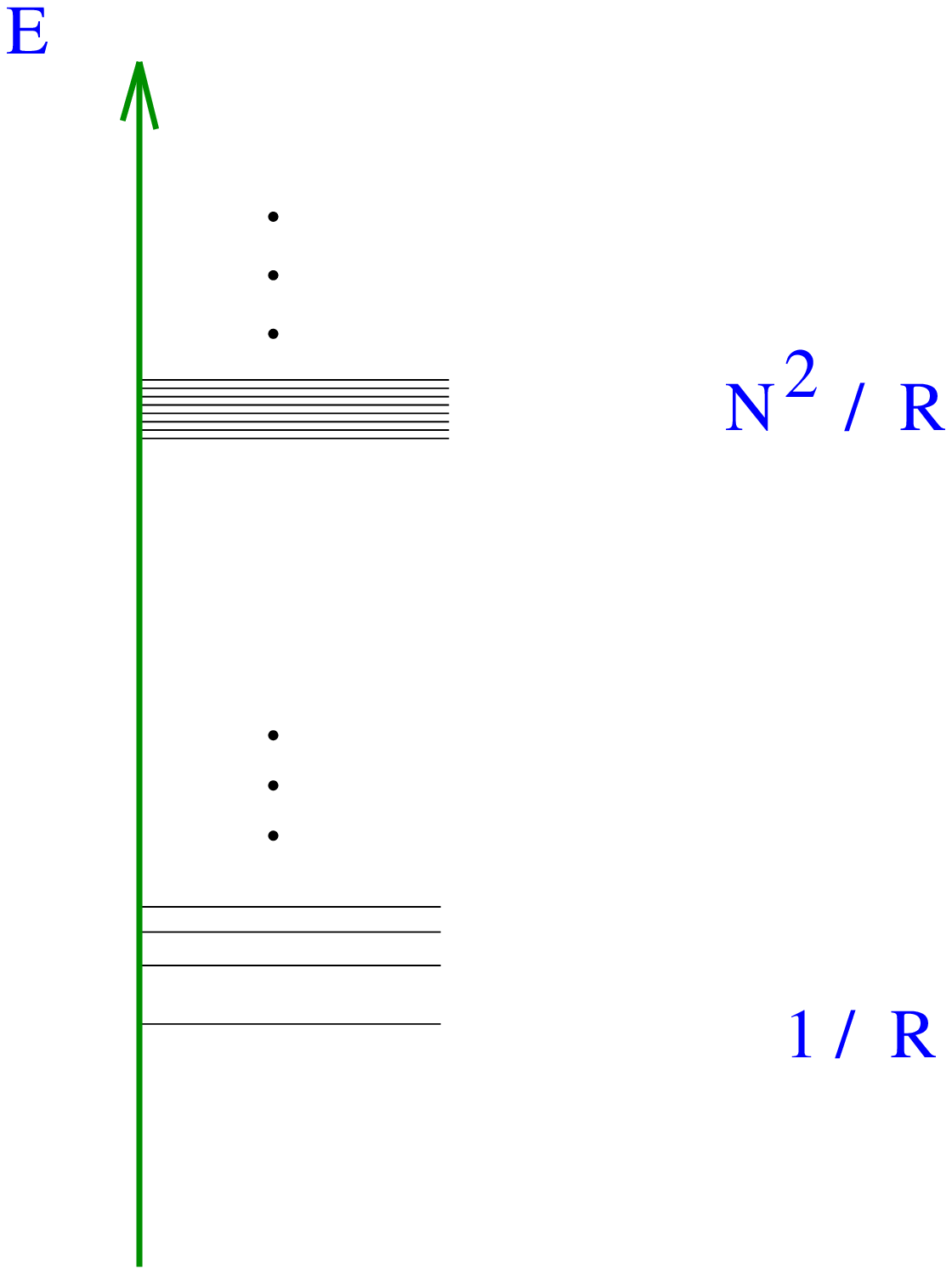}
\end{center}
%\vspace{1in}
\caption{
\small
\sl
 The energy spectrum of a CFT representing ${\rm AdS}_d$ quantum gravity. The
spectrum is discrete on a sphere of radius $R$, with gap of order $1/R$.
 The asymptotic energy band of
very dense ``black hole" states
  sets in beyond energies of order $N^2 /R$. The corresponding density of states is
that of a conformal fixed point in $d-1$ spacetime dimensions.}
\label{figure1}
\end{figure}

For large eternal AdS black holes with Schwarschild radius $R_S \gg R$
one may attempt to rise to the challenge, since they 
are thermodynamically stable and can exist in equilibrium
with thermal radiation at fixed  (high) temperatures $ 1/\beta \gg 1/R$.
Indeed, the corresponding Bekenstein--Hawking entropy
scales like that of $N^2$ conformal degrees of freedom at high energy,
\beq\label{dens} 
S \sim N^{2\over d-1} \;(E\,R)^{d-2 \over d-1} \sim N^2 \, (R/\beta)^{d-2} \;.
\eeq
Therefore, large AdS black holes with large Hawking temperature
$\beta^{-1} \gg R$ describe
the leading approximation to the thermodynamical functions of the canonical CFT state
\beq\label{cano} 
\rho_\beta =
 {e^{-\beta H} \over Z(\beta)}\;, \qquad Z(\beta) = {\rm Tr} \,\exp(-\beta H)\;.
\eeq

This suggests that we can test the semiclassical unitarity argument by careful analysis
of slight departures from thermal 
equilibrium, rather than studying a complete evaporation instability
in the vacuum.
Ref. \cite{rmaldas} proposes to look at the very long time structure of
 correlators of the form
\beq\label{timec} 
G (t) = {\rm Tr}  \,\left[\,\rho\,A(t)\,A(0)\,\right]\;,
\eeq
for appropriate Hermitian operators $A$. In this expression, $\rho$ denotes
the density matrix of the initial state. 
In the semiclassical approximation, one expects
these correlators to decay as $\exp(-\Gamma \,t)$
 with $\Gamma \sim \beta^{-1}$. However,
because the CFT lives in finite volume,
 the spectrum is actually discrete (c.f. Fig 1), and the
correlator must show nontrivial long time 
structure in the form of Poincar\'e recurrences, in particular it does
not vanish 
 (see \cite{rsuscumple, rsuspodos}).
This result,
which is straightforward from the boundary theory point of view,
has far reaching
consequences as far as the bulk physics is concerned.

\begin{figure}[hp] 
%\hspace*{0.1in}
\begin{center} 
\epsfysize=2.5in
\epsffile{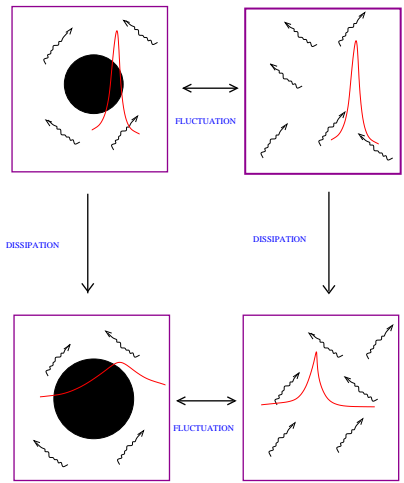}
\end{center}
%\vspace{1in}
\caption{
\small
\sl
A detailed analysis of dissipation of fluctuations in a finite thermal system
can reveal the effect of large quantum fluctuations in which a black hole 
turns into thermal radiation and viceversa. In the semiclassical approximation
to quantum gravity, these processes are represented by a coherent sum
over saddle points of different topology. In the case at hand we can use AdS
space as an effective finite-volume box.}
\label{figure2}
\end{figure}

Hence, the non-vanishing of $G(t)$  as $t\rightarrow \infty$ can be used as a criterion
for unitarity preservation
 beyond the semiclassical approximation. This argument can be made more
explicit by checking the effect of coherence loss on the long-time behaviour of $G(t)$.
Using the results of \cite{rpesk} one can simulate the required decoherence by coupling
an ordinary quantum mechanical system to a random classical noise. It is then shown in
\cite{rsus} that this random noise forces $G(t)$ to decay exponentially
for large $t$, despite having a discrete energy spectrum. This indicates
that the long-time
behaviour of correlators probes the strict quantum coherence of the bounded system.

At the same time, one would like to identify what kind of systematic corrections to the
leading semiclassical approximation are capable of restoring unitarity. A proposal was made
in \cite{rmaldas} in terms of topology-changing fluctuations of the AdS background. Our
purpose here is to investigate these questions and offer an explicit estimate of the
instanton effects suggested in \cite{rmaldas}. We conclude  (c.f. \cite{
rsus}) that large topological fluctuations are unlikely to restore unitarity
in full detail, although they represent a step forward. In particular, 
certain coarse-grained properties, such as the time averages of the correlators
(\ref{timec}), are   reproduced in order of magnitude. Related
work, especially in the $d=3$ case, can be found in refs. \cite{rsolo, raul,
solo}.

These considerations should also shed light on the recent proposal in  
\cite{rhawtalk}, where topological diversity is credited with the restoration of
 S-matrix unitarity in
black hole formation and evaporation.

\section{Long-time details of thermal quasi-equilibrium}

\noindent

Poincar\'e recurrences occur in general bounded systems.
 Classically they follow from
the compactness of available phase space, plus  the preservation of
the phase-space volume in time (Liouville's theorem).
 Quantum mechanically, they follow from discreteness of the energy spectrum
(characteristic of spatially bounded systems) and unitarity. The correlator
(\ref{timec}) 
\beq\label{quasi} 
G (t) = 
 \sum_{j,k} C_{jk}\,\;e^{i(E_j -E_k)t}
\;,\qquad {\rm with}\;\;\;\;C_{jk} = \sum_{i} \rho_{ij} A_{ki} A_{jk}
\;,\eeq 
defines a quasiperiodic function of time, provided the matrix elements
$C_{ij}$ are sufficiently bounded so that  $G(t)$ is well defined.  
 After initial dissipation
on a non-universal 
 time scale $\Gamma^{-1}$, where $\Gamma$ measures the approximate width of 
the matrix  elements $C_{ij}$  
in the energy basis,
the correlator will show large ``resurgences" when most of the relevant phases complete
a whole period (c.f. Fig 3). The associated time scale is
 $t_H \equiv 1/\langle \omega \rangle$, with
$\langle \omega \rangle = \langle E_i - E_j \rangle$ an average frequency in 
(\ref{quasi}). We can estimate
$\langle \omega \rangle$ as $\Gamma /\Delta n_\Gamma$, where $\Delta n_\Gamma$ is the number of energy
levels in the relevant band of width $\Gamma$. This must be proportional to
the density of levels, or the exponential of the entropy, i.e. we have  
\beq\label{thes}
t_H \sim \Gamma^{-1}  \;e^{S(\beta)}\,.
\eeq

\begin{figure}
%\hspace*{1.4in}
\center
\epsfysize=2.0in
\epsffile{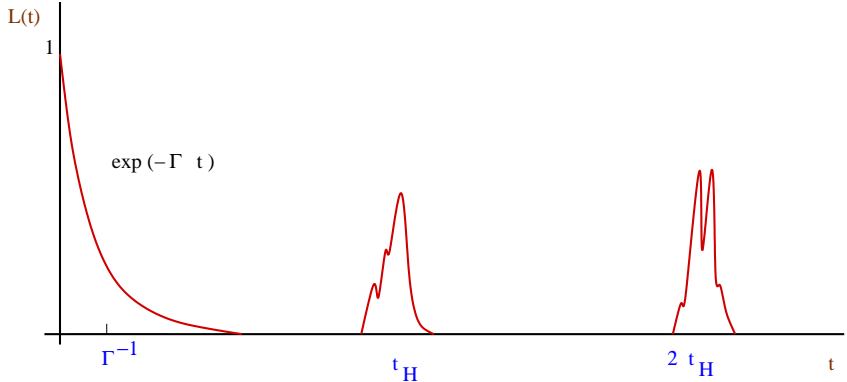}
%\vspace{1in}
\caption{
\small
\sl
 Schematic representation of
the  very long time behaviour of the normalized time correlator $L(t)$ in bounded systems.
The initial decay with lifetime of order $ \Gamma^{-1}$
is followed by O(1) ``resurgences" after the Heisenberg time $t_H \sim 
\Gamma^{-1}\,\exp(S)$
has elapsed. Poincar\'e recurrence times can be defined by demanding the resurgences to
approach unity with a given {\it a priori} accuracy, and  scale like
a double exponential of the entropy.
}
\label{figure3} 
\end{figure}

Following \cite{rsrednicki} we call this the   Heisenberg time scale.
 Poincar\'e times
can be defined in terms of quasiperiodic returns of $G (t)$
 with a given {\it a priori} accuracy. In a
sense, the Heisenberg time is the smallest possible Poincar\'e time.

A more quantitative, albeit more inclusive criterion
 can be used by defining a normalized positive correlator,
$L(t)$, satisfying $L(0)=1$, and its infinite time average,
\beq\label{eledef} 
L(t) \equiv \left|{G(t) \over G(0)}\right|^2, \qquad {\overline L} \equiv \lim_{T\rightarrow
\infty} {1\over T} \int_0^T dt \,L(t)\;.
\eeq 
The profile of $L(t)$ is sketched in Fig 3. The time average can be estimated by noticing
that the graph of $L(t)$ features positive ``bumps" of height $\Delta L$ and width $\Gamma$,
separated a time $t_H$, so that
\beq\label{fores} 
{\overline L} \sim {\Delta L \over  \Gamma \,t_H}\;.
\eeq 
For the case at hand $\Delta L \sim 1$,  $t_H \sim \Gamma^{-1} \,e^S$, 
and we obtain
 (c.f. \cite{rsuspodos,rsus})
\beq\label{otr} 
{\overline L} \sim \exp\left(-S(\beta)\right) \;.
\eeq 
Hence, the ``recurrence index"  scales as ${\overline L} \sim \exp(-N^2)$ in the
high-temperature phase. Since $N^2 \sim G_{\rm N}^{-1}$ in the AdS/CFT
dictionary, the index scales  as a nonperturbative
effect in the semiclassical approximation.

\section{Absence of recurrences in semiclassical black holes}

\noindent 

The previous considerations suggest that recurrences should be invisible in
gravity perturbation theory, i.e. in an expansion in powers of $1/N^2$ 
 around a black hole solution, 
  and this is
indeed what is found. 
The reason is that the relevant eigenfrequencies $\omega$ (the so-called
normal modes of the black hole) form a continuous spectrum
to all orders in the $1/N$ expansion. 
 For a static metric
 of the form
\beq\label{bhk} 
ds^2 = -g(r)\,dt^2 + {dr^2 \over g(r)} + r^2 \,d\Omega_{d-2}^2 \;,
\eeq 
the normal  frequency spectrum follows from the diagonalization of a radial Schr\"odinger operator
 \beq\label{veun}
 \omega^2 = - {d^2 \over dr_*^2} +
 V_{\rm eff} (r_*)
   \;,
\eeq 
   with
   \beq\label{vedo} 
   V_{\rm eff} = {d-2 \over 2}\,g(r)\left({g'(r)\over r}  +
   {d-4 \over 2r^2} \,g(r) \right) + g(r)  \left(-{\nabla^2_\Omega
   \over r^2} + m^2 \right)
   \;
\eeq 
   for a scalar field of mass $m$ (analogous potentials can be deduced
for higher spin fields). Here we have defined  the
 Regge--Wheeler or   ``tortoise" coordinate
   $ dr_* = dr / g(r)
  $.

\begin{figure}
%\hspace*{1.4in}
\center
\epsfysize=1.5in
\epsffile{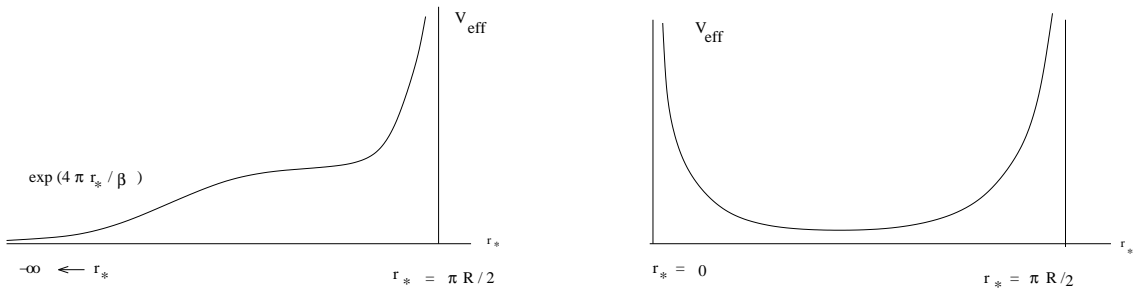}
%\vspace{1in}
\caption{
\small
\sl
 The effective potential determining the semiclassical
 normal frequency modes in a large AdS black hole
background (left).
 In Regge--Wheeler coordinates the horizon is at $r_* = -\infty$, whereas
the boundary of AdS is at $r_* = \pi R/2$ (only the region exterior to the horizon
appears). There is a universal exponential behaviour
in the near-horizon (Rindler) region. The effective one-dimensional Schr\"odinger problem
 represents a semi-infinite
barrier 
with a  continuous energy spectrum. This contrasts with
the analogous effective potential in vacuum AdS with global coordinates (right). The
domain of $r_*$ is  compact and the spectrum of normal modes is discrete with gap of
order $1/R$.}
\label{figure4}
\end{figure}

We have shown in Fig. 4 the form of the resulting effective potentials for large AdS black
holes, compared with the case of the vacuum AdS manifold.
The vacuum AdS manifold, corresponding to the choice
$g(r) = 1+r^2 /R^2$ in (\ref{bhk}),
behaves like a finite cavity, as expected.  The distinguishing feature
of the black-hole  horizon is a
 a non-degenerate
zero, $g(r_0) =0$, which induces   the universal scaling
\beq\label{univs}
 V_{\rm eff} (r_*) \;\propto \;\exp(4\pi r_* /\beta) \;\;\;\;{\rm as}\;\;\;
 r_*
\rightarrow -\infty\;, 
\eeq
 with $1/\beta
= g'(r_0) /4\pi$ the Hawking temperature and the horizon $r=r_0$ appearing
at $r_* = -\infty$. Notice that the near-horizon behaviour (\ref{univs}) only
depends on the Hawking temperature, i.e. the curvature at the horizon, and
is independent of other long-distance features of the gravitational background.
      The
 spectrum is  discrete in pure AdS, and continuous in the AdS black hole.
Physically, this just reflects the fact that the horizon is an infinite redshift surface, so
that we can store an arbitrary number of modes with finite total energy, provided they are
sufficiently red-shifted by approaching the horizon \cite{rbrick}.
Since the thermal entropy of perturbative gravity excitations
 in the vacuum AdS spacetime scales as
$S(\beta)_{\rm AdS} \sim N^0$, we see that the perturbative Heisenberg time of
the AdS spacetime is of $O(1)$ in the large-$N$ limit, leading to ${\overline L}_{\rm AdS}
= O(1)$. On the other hand, we have ${\overline L}_{\rm bh} =0$ in this approximation.

Although these
results are based on the leading perturbative approximation in the
classical black hole background, it is unlikely that higher-order
perturbative effects will render the frequency spectrum discrete, because
this feature appears as an infrared property of the potentials in
Fig. 4 (c.f. \cite{barbonh}). 

Another argument for the robustness of ${\overline L}_{\rm bh}$ in perturbation
theory  comes from the 
Euclidean formalism, obtained by $t=-i\tau$ in (\ref{bhk}), followed by an identification
$\tau \equiv \tau + \beta$.  The resulting metric for the vacuum AdS spacetime
has a non-contractible ${\bf S}^1$ given by the $\tau$ compact direction. We call
$Y$ this Euclidean manifold. On the other hand, the black hole spacetime with
$g(r_0)=0$ is simply connected,  since the thermal ${\bf S}^1$ shrinks to zero
size at $r=r_0$. The choice $1/\beta = g'(r_0)/4\pi$ ensures smoothness at
$r=r_0$. We call this Euclidean black hole manifold $X$.

\begin{figure}
%\hspace*{1.4in}
\center
\epsfysize=2in
\epsffile{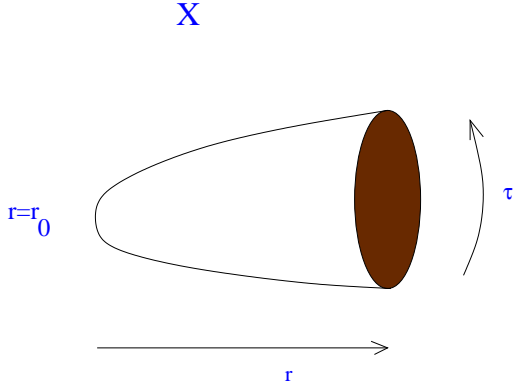}
%\vspace{1in}
\caption{
\small
\sl
The  Euclidean black hole manifold $X$ is
simply connected, unlike standard thermal manifolds in quantum field
theory.}  
\label{figure6}
 \end{figure}

The real-time correlation functions in the black hole background, $G(t)_X$,
 follow by analytic continuation from
their Euclidean counterparts. Since $X$ is a completely smooth manifold in the
$1/N$ expansion, so is the Euclidean correlator $G(it)_X$ for $t > 0$.
The continuous
spectrum arising in the spectral decomposition of $G(t)_X$ is a consequence of
the contractible topology of $X$, since the  foliation by $\tau={\rm constant}$
hypersurfaces is singular at the horizon 
 (c.f. \cite{rsus}). 

\begin{figure}
%\hspace*{1.4in}
\center
\epsfysize=2in
\epsffile{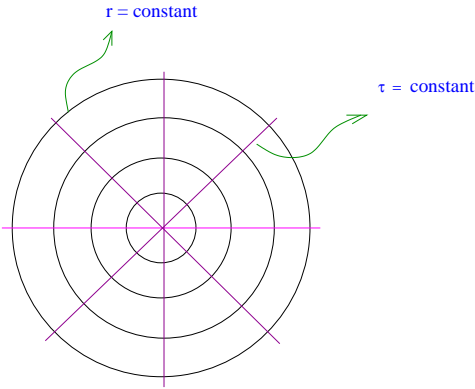}
%\vspace{1in}
\caption{
\small
\sl
In the vicinity of  $r=r_0$, the
manifold $X$ is well-approximated by the product of a flat disk and the
${\bf S}^{d-2}$ at the horizon.
Equal-time hypersurfaces of 
Hamiltonian foliations on $X$ have a fixed point at  $r=r_0$.
This fact is responsible for {\rm both} the classical contribution to
the entropy, and the continuous spectrum of normal frequencies.} 
\label{figure7}
 \end{figure}

Since the continuous spectrum finds its origin in the topological properties
of $X$, this particular fact will not be affected by perturbative corrections.
Incidentally, the same peculiar behaviour with respect to the Hamiltonian
conjugate to $\tau$ is responsible for the existence of a formally  classical
entropy. Namely, the Euclidean action
\beq
\label{eac}
I(X)= -{1\over 16\pi G_{\rm N}} \int_X \; {\cal R} -{1\over 8\pi G_{\rm N}}
\oint_{\partial X}
\;\left({\cal K} + {\rm C.T.}\right)
\eeq
with appropriately defined counterterms, ${\rm C.T.}$, is not just given by  
$\beta M_{\rm ADM}$,
despite the fact that $\partial_\tau$ is a Killing vector on $X$. 
Rather, one finds \cite{rgh} 
\beq
\label{een}
I(X)= \beta\,M(X) - S_{\rm bh} (X)
\;,
\eeq
with $S_{\rm bh} = A_{\rm H} /4G_{\rm N}$ the Bekenstein--Hawking entropy.  
The microscopic interpretation of this entropy must be referred back to
the dual CFT. This point of view suggests that the information encoded in the
geometry of $X$ is   fundamentaly 
coarse-grained, so that the continuous spectrum of
frequencies would also 
be a reflection of this coarse-graining.

\section{Topological diversity and unitarity}

\noindent

Our discussion in the previous section suggests that  improving on the
semiclassical prediction ${\overline L}_{\rm bh} =0$ requires some
sort of topology-change process. 
The proposal of \cite{rmaldas}
  is
precisely that: instead of evaluating the semiclassical correlators on $X$, one
 should
sum coherently the contribution of $X$ and $Y$. Normally one neglects the contribution of
$Y$ on a quantitative basis (at high temperatures $R\gg \beta$).
 However, here the contribution of $X$ to ${\overline L}$
vanishes and one is forced to consider the first correction.  Since $Y$ has a discrete spectrum
in perturbation theory, the net result for ${\overline L}$ should be non-vanishing in
this approximation.
Physically, this superposition of Euclidean saddle points (or master fields, in the
language of the CFT) corresponds to large-scale fluctuations in which the AdS
black hole is converted into a graviton gas at the same temperature and viceversa.

The resulting time profile in the instanton approximation takes the form 
(c.f. \cite{rsus})
\beq
\label{tprof}
L(t)_{\rm inst} = L(t)_X + C\,e^{-2\Delta I} \; L(t)_Y
\;
\eeq
where $C= O(N^0)$, $\Delta I = I_Y - I_X$ and $I=-\log \,Z(\beta)$, calculated in the classical
gravity approximation. Since $I_Y \sim -N^0$ and $I_X \sim -N^2$, the exponential
suppression factor is of order $\exp(-2 |I_X|) \sim \exp(-N^2)$. 
The resulting structure is shown in Fig 8. The instanton approximation to the
normalized correlator features the normal dissipation with lifetime $\Gamma^{-1} \sim
\beta$ coming from the contribution of $X$. However, the resurgences are controlled by
$L(t)_Y$,
 damped by a factor $\exp(-2\Delta I) \sim \exp(-N^2)$, and separated a time
$t_H (Y) \sim N^0$.

\begin{figure}
%\hspace*{1.4in}
\center
\epsfysize=2in
\epsffile{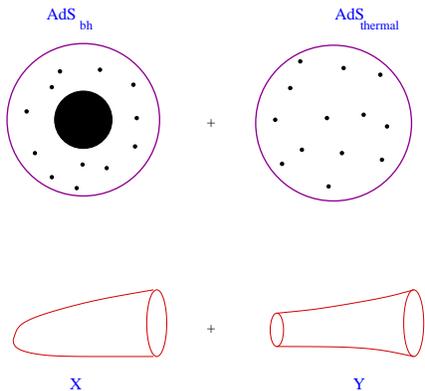}
%\vspace{1in}
\caption{
\small
\sl
 Summing over  large-scale fluctuations of the thermal ensemble in which
a black hole spontaneously turns into radiation (and viceversa) is represented
in
the
Euclidean formalism
as the coherent sum of thermal saddle points of different topology.
The ``cigar-like" geometry $X$
represents the black-hole master field (in the CFT
language) and the cylindrical topology $Y$ represents the thermal gas of particles.
}
\label{figure8}
\end{figure}

We can also find the time scale $t_c$ for which the large-scale instantons considered
here are quantitatively important
on the graph of $L(t)$. This is shown in Fig. 8 and
yields $t_c \sim \Delta I /\Gamma \sim N^2$.

We see that the instanton approximation does not reproduce the
expected pattern of recurrences, particularly at high temperatures. It
is interesting to find out how much this depends on the temperature above
the phase transition.

For large AdS black holes, positive specific heat sets in for $r_0 > R \left(
{d-3 \over d-1}\right)^{1/2}$, but these black holes do not dominate the 
leading large-$N$ thermodynamics until $r_0 = R$, the location of the
Hawking--Page phase transition \cite{rhpage}. In the immediate vicinity of the transition
the statistical weight of the two backgrounds is approximately the same,
since $\Delta I \approx 0$. However, the entropy increases by a factor
of order $N^2$ across the transition, and we would expect a sharp change
of behaviour of $t_H$ as a function of $\beta$, as well as the long-time
structure of $L(t)$.

On the other hand, in the instanton approximation the resurgences are
controlled by $t_H (Y)$, which is  of $O(1)$ in the large-$N$ limit on
both sides of the phase transition. The only difference is that $\Delta I$
starts increasing away from zero as the temperature increases. The bumps,
spaced $t_H (Y)$ appart and initially of height $O(1)$, 
decrease accordingly in size. When
$(r_0 -R)/R \sim 1/N^2$, we reach $\Delta I \sim N^2$ and the pattern 
in Fig. 8. 

\begin{figure}
%\hspace*{1.4in}
\center
\epsfysize=1.8in
\epsffile{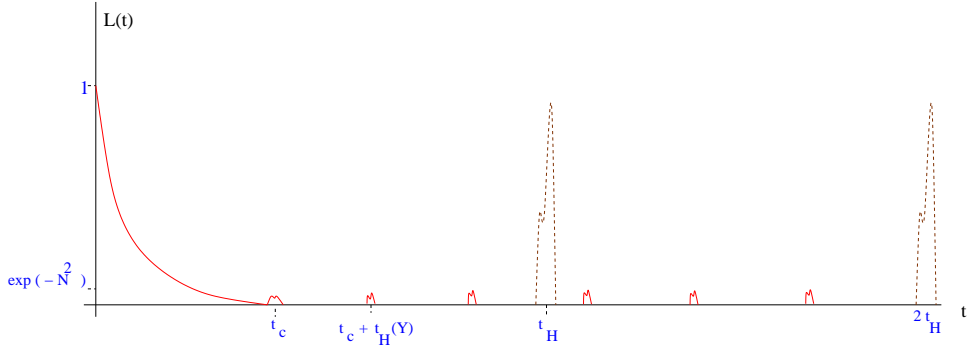}
%\vspace{1in}
\caption{
\small
\sl
The instanton approximation to the correlator $L(t)_{\rm inst}$ features
the expected exponential decay $\exp(-\Gamma \,t)$ induced by the contribution of t
he
$X$-manifold, whereas the resurgences
are entirely due to the  interference with the $Y$-manifold, leading to
small bumps of order $\exp(-2\Delta I) \sim \exp(-N^2)$, separated a time
$t_H (Y) \sim N^0$. These bumps are noticeable against the background of the $X$-ma
nifold
after a time $t_c \sim \Delta I /\Gamma$. In the dashed line we plot the very
different expected behaviour in the exact CFT: large $O(1)$ bumps separated
by time intervals of order $\exp(N^2)$. Despite the gross differences between
both profiles, their time averages coincide in order of magnitude. }
\label{figure5}
\end{figure}

In the limit of very high temperatures, there are some limitations to
be expected. The free energy of the $Y$ manifold scales as that
of a $D$-dimensional thermal gas, with $D=10$ or $D=11$ depending on
the particular model of AdS/CFT duality considered, i.e.
$I(Y) \sim - (R/\beta)^{D-1}$. Hence, at temperatures of order
\beq
\label{vlar}
R/\beta \sim N^{2 \over D-d-3}
\eeq
the $Y$ manifold would dominate again over $X$ (c.f. \cite{rexten}).
 However, perturbative
instabilities of $Y$ appear before this threshold. For example, in
the standard case of ${\rm AdS}_5 \times {\bf S}^5$ duality, the
$Y$ manifold reaches the Hagedorn instability at temperatures
$R/\beta \sim (g_s N)^{1/4}$, and the Jeans instability at temperatures
$R/\beta \sim N^{1/5}$.   
 
Despite all these caveats,  the instanton approximation
yields an interesting value for  
   the infinite time average, at {\it all} temperatures. 
\beq\label{isntl}
{\overline L}_{\rm inst} \approx C\;e^{-2\Delta I}\;
.
\eeq
Namely, 
we have
\beq\label{degg} 
{\overline L}_{\rm inst} \sim {\Delta L \over \Gamma \,t_H} \sim {e^{-N^2} \over \Gamma \cdot \Gamma^{-1}}
\sim {1 \over  \Gamma \cdot \Gamma^{-1} \,e^{N^2}} \sim {\overline L}_{\rm CFT}\;.
\eeq 
The first estimate obtains ${\overline L}_{\rm inst} \sim \exp(-N^2)$ from
the Boltzman suppresion of the $Y$ manifold, despite the fact that $t_H (Y) 
\sim O(1)$, whereas the second estimate is based on $O(1)$ recurrences with
very large Heisenberg time. It is important to stress that (\ref{degg}) holds
up to factors of order $\exp(-c N^2)$ with $c=O(1)$, because in general
$S_X \neq -2|I_X|$, even
at high temperatures \cite{rsus}. For large AdS black holes,
the instanton calculation (\ref{degg}) gives a larger value of the index
than the estimate based on the quantum mechanical density of states    
 (\ref{otr}). 

Thus, we find that a coarse-grained question, such as 
 the infinite-time averages of correlators, is better  accounted for
by the semiclassical instanton approximation than a ``detailed" question,
such as the concrete time structure of the correlators. This is
 another indication of
the fundamentally thermodynamical features of relativistic horizons.

A more complicated set of Euclidean saddle points can be analyzed for the
three-dimensional case of BTZ black holes. The authors of \cite{raul, solo}
 conclude
that resummation of an infinite family of  $SL(2,{\bf Z})$ saddle points
 is unlikely to alter the conclusions presented here on the basis of the leading
instanton approximation. The authors of
Ref.  \cite{raul} also  point out that only a finite
set of black-hole saddle points remains under the control of the semiclassical
approximation after a time of order $t_c \sim c$, where $c$ is the central
charge of the CFT. 
 
\begin{figure} 
%\hspace*{1.4in}
\center
\epsfysize=2in
\epsffile{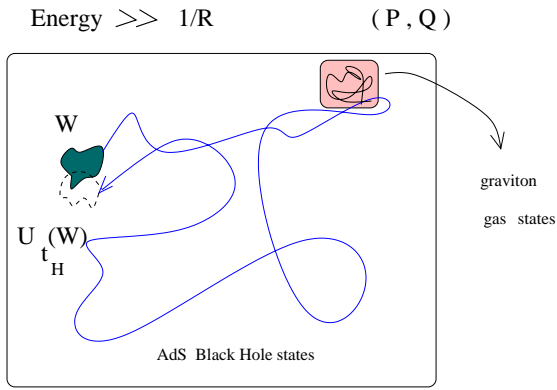}
%\vspace{1in}
\caption{
\small
\sl
A pictorial representation of the compact phase space at very large energy.
Poincar\'e recurrences consist on the time development $U_t (W)$ of a
region $W$ intersecting itself after a period larger than  $t_H$. We have
separated the dominant black-hole like states from the relatively scarce
thermal gas states in the phase space. The instanton approximation is
only sensitive to the recurrences in the small patch of thermal gas states,
because the spectrum
of black-hole states is effectively treated as   continuous.
}
\label{figure9}
\end{figure}

\section{Conclusions}

\noindent

The study of  very long time features of correlators in black hole backgrounds is
a potentially important approach towards unraveling the mysteries of black hole
evaporation and the associated physics at the spacelike singularity.
We have seen that large scale topology-changing fluctuations proposed in
 \cite{rmaldas} 
begin to restore some of the fine structure required by unitarity, but fall short at
the quantitative level. Presumably the appropriate instantons occur on 
microscopic
 scales and
involve stringy dynamics.

 While  semiclassical black holes do
 faithfully reproduce
``coarse grained" inclusive properties of the system such as the entropy
(c.f. \cite{rgh}),  additional dynamical
features of the horizon may be necessary to resolve finer details of the
information
loss problem.
Roughly, one needs a systematic set of corrections that could
generate a ``stretched horizon" of Planckian thickness \cite{rstre, rbhc}.
 The crudest model of such
stretched horizon is the brick-wall model of 't Hooft \cite{rbrick}. In this phenomenological
description  one replaces the horizon by a reflecting boundary condition at Planck distance
$\epsilon \sim \ell_P$ 
from the horizon. This defines a ``mutilated" $X_\epsilon$ manifold, of cylindrical
topology,
 leading to a discrete spectrum of the right spacing in order of magnitude.
Notice that, in line with our previous discussion, the discrete spectrum on
the effective manifold $X_\epsilon$ is tied with the absence of classical
contribution to the entropy, whose leading contribution is obtained at
one loop order: $S(X_\epsilon) \sim A_{\rm H} / \epsilon^{d-2}$. 

We have also seen that the characteristic time for large topological fluctuations to be
important is $t_c \sim O(N^2)$ in the semiclassical approximation. In 
\cite{rshenk}  it
was argued that semiclassical two-point functions  probe  the black hole singularity
on much shorter characteristic times, thereby justifying the analysis on the single standard
black hole manifold.  However,  detailed
 unitarity is only  restored on
time scales of order $t_H \sim \exp(N^2)$. Thus $t_c \ll t_H$ and we conclude that such
semiclassical analysis of the singularity is bound to be incomplete, as it misses
 whatever
microphysics is responsible for the detailed
 unitarity restoration in the quantum mechanical
time evolution.

\begin{figure}
%\hspace*{0.1in}
\center
\epsfysize=2in
\epsffile{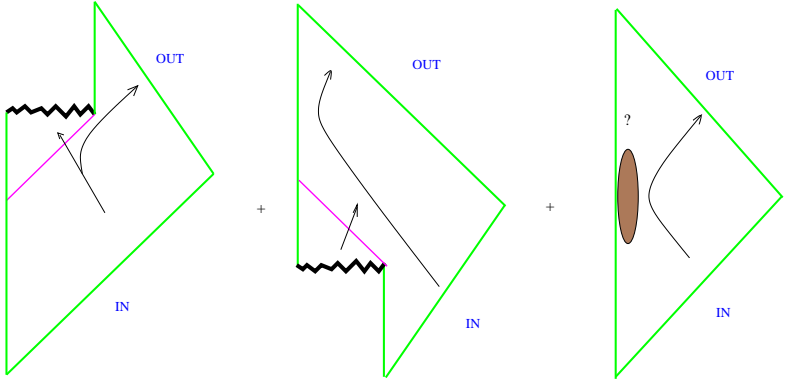}
%\vspace{1in}
\caption{
\small
\sl
Different topological contributions to the path integral of the scattering
matrix.  We have drawn classical black hole and white hole spacetimes (for
CPT invariance), as well
as a spacetime of trivial topology. According to \cite{rhawtalk},
trivial topology
contributions would be enough to restore unitarity of the S-matrix.}
\label{haw}
\end{figure}

It is natural to ask at this point what  possible  lessons can be drawn  
regarding the related problem  of the black hole S-matrix. In particular,
Ref.  \cite{rhawtalk} uses the 
 main idea of \cite{rmaldas}, extrapolating it to
the S-matrix problem, and claiming that trivial-topology spacetimes contributing
to the path integral are enough to restore unitarity in the complete quantum
amplitudes (c.f. Fig. 10).    

The standard black hole spacetimes in Fig. 10 violate quantum coherence
because either the {\it In} or {\it Out}
 asymptotic surfaces do not support the complete
{\it In} or {
\it Out} Hilbert space, whereas the topologically trivial spacetimes do
have complete {\it In} and {\it Out} asymptotic surfaces. Thus, the claim
would be that asymptotic completeness for off-shell histories is  
enough to cure the lack of asymptotic completeness of many approximately on-shell
histories. Since asymptotic completeness is the key property guaranteeing   
S-matrix unitarity (c.f. \cite{rlag, rpesk}), it is hard to imagine  that
a path integral featuring {\it both} topologically trivial {\it and}
topologically nontrivial spacetimes could be {\it exactly} unitary. A
more natural expectation would be that such topological diversity could
at best achieve some approximate restoration of unitarity.    

One  obstacle in assesing the proposal of \cite{rhawtalk} is the technical
difficulty in estimating the quantitative effect of trivial topologies
in the complete path integral. For initial conditions that would
classically produce a black hole, the spacetimes of trivial topology
are far from any semiclassical saddle point, and therefore their contribution 
is difficult to evaluate with the required precision. In fact, the 
existence of a classical saddle point
with trivial topology, the manifold $Y$, was the main reward for
choosing thermal equilibrium states in Ref. \cite{rmaldas}, as opposed to
S-matrix boundary conditions. 

The situation with the topologically trivial spacetimes in the S-matrix
is perhaps analogous to that of the $Y$ manifold at temperatures above
the Jeans instability, when it becomes unstable and hence ceases to be a
good saddle point of the path integral.
 At sufficiently high temperatures, the Jeans length
$\ell_J \sim \sqrt{\beta^{d}/G_{\rm N}}$ falls below the curvature radius of
AdS, and some thermal fluctuations of wavelength  $\lambda > \ell_J$
develop an   imaginary effective  mass. In real time, this corresponds to
exponential behaviour of linearized perturbations, proportional to
$\exp(\pm t/\ell_J)$, thus entering the non-linear regime beyond the
applicability of 
perturbation theory. On physical grounds, we expect that the endpoint
of this ``tachyonic instability" is the large AdS black hole at
the corresponding temperature, thus reverting back to the $X$ manifold
as the unique stable saddle point. 

\begin{figure}
%\hspace*{0.1in}
\center
\epsfysize=2in
\epsffile{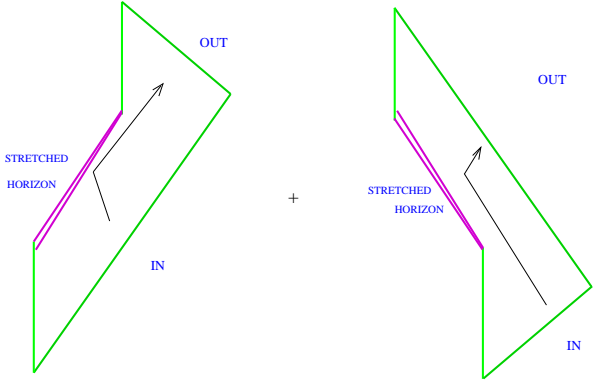}
%\vspace{1in}
\caption{
\small
\sl
Topologically trivial spacetime describing unitary scattering according
to the principle of black hole complementarity. The picture is not
fully semiclassical because the stretched horizon has no known
semiclassical description, except for thermodynamical, coarse-grained 
observables. The ``white hole" stretched horizon geometry has been
introduced in order to respect CPT invariance. }
\label{hawt}
\end{figure}

Therefore, if we are to draw inspiration from the study of thermal boundary
conditions, we would suggest that topological diversity {\it per se}
is not enough to restore unitarity of the S-matrix, unless some 
coarse-graining is imposed on the S-matrix itself (the analog of 
calculating of ${\overline L}$ as opposed to $L(t)$.) 

On general grounds, the AdS/CFT correspondence
suggests that spacetime topology can
be unambiguously defined only in the context of the semiclassical
approximation. Perhaps the best we can do in representing geometrically
the black-hole
S-matrix is the effective spacetime suggested by the principle of
black hole complementarity (c.f. \cite{rbhc}), which involves a topologically
trivial spacetime with a stretched horizon that behaves as a
boundary equipped with an effective Hilbert space and effective Hamiltonian
yet to be found (c.f. Fig. 11). In this effective description of the
 stretched horizon, all the quantum fluctuations would be included, and
thus no further summation over topological sectors would be needed.

Perhaps the following metaphor will turn out to be a useful guide line.
In massless QCD an infrared scale is required  to control the large
perturbative infrared fluctuations. The  scale is indeed dynamically generated  
in asymptotically free systems. Although gravity is {\it a priori} equipped with an
 intrinsic
Planck  scale,  
in the near-horizon region this scale 
is red-shifted away and
the system seems to posses no scale (c.f. the universal potential of
(\ref{univs})). A dynamically produced scale would presumably 
allow the formation of a streched horizon and the restoration unitarity.
Some universal features of such stretched horizon were recently pointed out
in \cite{thor}.

%\vspace{0.2cm}
%%%%%%%%%%%%%%%%%%%%%%%%%%%%%%%%%%%%%%%%%%% 

\section*{Acknowledgements}

 E. R. would like to thank the KITP at Santa Barbara
for hospitality during the completion of this work, under grant of the  National Science
Foundation  No. PHY99-07949.
 The work of J.L.F.B. was partially supported by MCyT
 and FEDER under grant
BFM2002-03881 and
 the European RTN network
 HPRN-CT-2002-00325. The work of E.R. is supported in part by the
BSF-American Israeli Bi-National Science Foundation, The Israel Science
Foundation
 and the European RTN network MRTN-CT-2004-512194.

%%%%%%%%%%%%%%%%%%%%%%

\end{document}